\documentclass[nofootinbib, preprintnumbers, showpacs, preprint]{revtex4-1}
\usepackage{amsmath, amssymb, braket}
\usepackage[dvipdfmx]{graphicx}
 \usepackage{epsfig}
 \usepackage{ulem}

\usepackage{color}

\newcounter{multieqs}

\newcommand{\be}{\begin{equation}}
\newcommand{\ee}{\end{equation}}
\newcommand{\eq}[1]{(\ref{#1})}
\newcommand{\bit}{\begin{itemize}}  \newcommand{\eit}{\end{itemize}}
\newcommand{\ben}{\begin{enumerate}}  \newcommand{\een}{\end{enumerate}}
\newcommand{\rf}[1]{(\ref{#1})}

\def\bd{\begin{document}}
\def\ed{\end{document}}
\def\bea{\begin{eqnarray}}
\def\eea{\end{eqnarray}}

\def\la{\langle}
\def\ra{\rangle}

\def\npb#1#2#3{Nucl. Phys. {\bf{B#1}} #3 (#2)}
\def\plb#1#2#3{Phys. Lett. {\bf{#1B}} #3 (#2)}
\def\prl#1#2#3{Phys. Rev. Lett. {\bf{#1}} #3 (#2)}
\def\prd#1#2#3{Phys. Rev. {D bf{#1}} #3 (#2)}
\def\cmp#1#2#3{Comm. Math. Phys. {\bf{#1}} #3 (#2)}
\def\cqg#1#2#3{Class. Quantum Grav. {\bf{#1}} #3 (#2)}
\def\nppsa#1#2#3{Nucl. Phys. B (Proc. Suppl.) {\bf{#1A}}#3 (#2)}
\def\ap#1#2#3{Ann. of Phys. {\bf{#1}} #3 (#2)}
\def\ijmp#1#2#3{Int. J. Mod. Phys. {\bf{A#1}} #3 (#2)}
\def\rmp#1#2#3{Rev. Mod. Phys. {\bf{#1}} #3 (#2)}
\def\mpla#1#2#3{Mod. Phys. Lett. {\bf A#1} #3 (#2)}
\def\jhep#1#2#3{J. High Energy Phys. {\bf #1} #3 (#2)}
\def\atmp#1#2#3{Adv. Theor. Math. Phys. {\bf #1} #3 (#2)}

\def\N{{\cal N}}
\def\sst{\scriptscriptstyle}
\def\thetabar{\bar\theta}
\def\Tr{{\rm Tr}}
\def\one{\mbox{1 \kern-.59em {\rm l}}}

\def\a{\alpha}      \def\da{{\dot\alpha}}  \def\dA{{\dot A}}
\def\b{\beta}       \def\db{{\dot\beta}}
\def\g{\gamma}  \def\G{\Gamma}  \def\dc{{\dot\gamma}}
\def\d{\delta}  \def\D{\Delta}  \def\ddt{\dot\delta}
\def\pd{{\dot\phi}}
\def\e{\epsilon}
\def\ve{\varepsilon}
\def\uve{\upvarepsilon}
\def\f{\phi}    \def\F{\Phi}    \def\vvf{\f}
\def\vphi{\varphi}
\def\h{\eta}
\def\k{\kappa}
\def\l{\lambda} \def\L{\Lambda}
\def\m{\mu} \def\n{\nu}
\def\o{\omega}
\def\p{\pi} \def\P{\Pi}
\def\r{\rho}
\def\s{\sigma}  \def\S{\Sigma}
\def\t{\tau}
\def\th{\theta} \def\Th{\Theta} \def\vth{\vartheta}
\def\X{\Xeta}
\def\z{\zeta}

\def\na{\nabla}

\def\cA{{\mathscr A}} \def\cB{{\cal B}} \def\cC{{\cal C}}
\def\cD{{\cal D}} \def\cE{{\cal E}} \def\cF{{\cal F}}
\def\cG{{\cal G}} \def\cH{{\cal H}} \def\cI{{\cal I}}
\def\cJ{{\mathscr J}} \def\cK{{\cal K}} \def\cL{{\cal L}}
\def\cM{{\cal M}} \def\cN{{\cal N}} \def\cO{{\cal O}}
\def\cP{{\cal P}} \def\cQ{{\cal Q}} \def\cR{{\cal R}}
\def\cS{{\cal S}} \def\cT{{\cal T}} \def\cU{{\cal U}}
\def\cV{{\cal V}} \def\cW{{\cal W}} \def\cX{{\cal X}}
\def\cY{{\cal Y}} \def\cZ{{\cal Z}}

\def\ua{\underline{\alpha}}
\def\uc{\underline{\phantom{\alpha}}\!\!\!\gamma}
\def\um{\underline{\mu}}
\def\ud{\underline\delta}
\def\ue{\underline\epsilon}
\def\una{\underline a}\def\unA{\underline A}
\def\unb{\underline b}\def\unB{\underline B}
\def\unc{\underline c}\def\unC{\underline C}
\def\und{\underline d}\def\unD{\underline D}
\def\une{\underline e}\def\unE{\underline E}
\def\unf{\underline{\phantom{e}}\!\!\!\! f}\def\unF{\underline F}
\def\unm{\underline m}\def\unM{{\underline M}}
\def\unn{\underline n}\def\unN{{\underline N}}
\def\unp{\underline{\phantom{a}}\!\!\! p}\def\unP{\underline P}
\def\unq{\underline{\phantom{a}}\!\!\! q}
\def\unQ{\underline{\phantom{A}}\!\!\!\! Q}
\def\unH{\underline{H}}

\def\As {{A \hspace{-6.4pt} \slash}\;}
\def\bs {{b \hspace{-6.4pt} \slash}\;}
\def\Ds {{D \hspace{-6.4pt} \slash}\;}
\def\Gts {{\Gt \hspace{-6.4pt} \slash}\;}
\def\ds {{\del \hspace{-6.4pt} \slash}\;}
\def\ss {{\s \hspace{-6.4pt} \slash}\;}
\def\ks {{ k \hspace{-6.4pt} \slash}\;}
\def\ps {{p \hspace{-6.4pt} \slash}\;}
\def\xs {{x \hspace{-6.4pt} \slash}\;}
\def\pas {{{p_1} \hspace{-6.4pt} \slash}\;}
\def\pbs {{{p_2} \hspace{-6.4pt} \slash}\;}
\def\cFs {{{\cal F} \hspace{-6.4pt} \slash}\;}
\def\Dss {{D \hspace{-7.5pt} \slash}\;}
\def\dss {{\del \hspace{-7.0pt} \slash}\;}

\def\Ah{{\hat{A}}}
\def\Dh{{\hat{D}}}
\def\Gh{{\hat{G}}}
\def\Fh{{\hat{F}}}
\def\Ih{{\hat{I}}}
\def\Jh{{\hat{J}}}
\def\Kh{{\hat{K}}}
\def\Lh{{\hat{L}}}
\def\Ph{{\hat{P}}}
\def\Rh{{\hat{R}}}
\def\Vh{{\hat{V}}}
\def\Xh{{\hat{X}}}

\def\ah{{\hat{\a}}}
\def\bh{{\hat{\b}}}
\def\gh{{\hat{\g}}}
\def\dh{{\hat{\d}}}
\def\rh{{\hat{\r}}}
\def\hh{\hat{h}}
\def\uh{\hat{u}}
\def\xh{\hat{x}}
\def\yh{\hat{y}}
\def\ph{\hat{p}}
\def\xih{\hat{\xi}}
\def\chih{\hat{\chi}}
\def\Psih{\hat{\Psi}}
\def\phih{\hat{\phi}}

\def\psit{\tilde{\psi}}
\def\Psit{\tilde{\Psi}}
\def\Psibt{\tilde{\bar{Psi}}}

\def\st{\tilde{\sigma}}

\def\delt{\tilde{\delta}}
\def\Phit{\tilde{\Phi}}
\def\Phitb{\overline{\tilde{Phi}}}
\def\tht{\tilde{\th}}
\def\lt{\tilde{\l}}
\def\chit{\tilde{\chi}}
\def\phit{\tilde{\phi}}

\def\At{\tilde{A}}
\def\Bt{\tilde{B}}
\def\Ct{\tilde{C}}
\def\Dt{\tilde{D}}
\def\Et{\tilde{E}}
\def\Ft{\tilde{F}}
\def\Gt{\tilde{G}}
\def\Ht{\tilde{H}}
\def\It{\tilde{I}}
\def\Jt{\tilde{J}}
\def\Qt{\tilde{Q}}
\def\Rt{\tilde{R}}
\def\Mt{\tilde{M }}
\def\Nt{\tilde{N}}
\def\St{\tilde{S}}
\def\Vt{\tilde{V}}
\def\Xt{\tilde{X}}
\def\at{\tilde{a}}
\def\ct{\tilde{c}}
\def\dt{\tilde{d}}
\def\htt{\tilde{h}}
\def\ft{\tilde{f}}
\def\gt{\tilde{g}}
\def\pt{\tilde{p}}
\def\qt{\tilde{q}}
\def\vt{\tilde{v}}
\def\nt{\tilde{n}}
\def\ut{\tilde{u}}
\def\wt{\tilde{w}}
\def\zt{\tilde{z}}
\def\xt{\tilde{x}}
\def\yt{\tilde{y}}
\def\Psit{\tilde{\Psi}}
\def\vphit{\tilde{\varphi}}
\def\tD{\tilde{\D}}
\def\tR{\tilde{R}}

\def\eb{\bar{\epsilon}}
\def\delb{\bar{\partial}}
\def\thb{\bar{\theta}}
\def\mub{\bar{\mu}}
\def\lamb{\bar{\l}}
\def\psib{\bar{\psi}}
\def\sb{\bar{\sigma}}
\def\xib{\bar{\xi}}
\def\chib{\bar{\chi}}

\def\Psib{\bar{\Psi}}
\def\Phib{\bar{\Phi}}
\def\Lamb{\bar{\Lambda}}
\def\Sb{{\overline \Sigma}}
\def\cb{\bar{c}}
\def\hb{\bar{h}}
\def\qb{\bar{q}}
\def\wb{\bar{w}}
\def\ub{\bar{u}}
\def\zb{{\bar{z}}}
\def\Hb{\bar{H}}
\def\Qb{{\bar Q}}
\def\Omegab{\overline{\Omega}}
\def\ob{\overline{\omega}}

\def\Ab{{\overline A}} \def\Bb{{\overline B}} \def\Cb{{\overline C}}
\def\Db{{\overline D}} \def\Eb{{\overline E}} \def\Fb{{\overline F}}
\def\Gb{{\overline G}}
\def\Ib{{\overline I}}
\def\Jb{{\overline J}} \def\Kb{{\overline K}} \def\Lb{{\overline L}}
\def\Mb{{\overline M}} \def\Nb{{\overline N}} \def\Ob{{\overline O}}
\def\Pb{{\overline P}}  \def\Rb{{\overline R}}
 \def\Tb{{\overline T}} \def\Ub{{\overline U}}
\def\Vb{{\overline V}} \def\Wb{{\overline W}} \def\Xb{{\overline X}}
\def\Yb{{\overline Y}} \def\Zb{{\overline Z}}

\def\fb{{\overline f}}
\def\gb{{\overline g}}
\def\mb{{\overline m}}
\def\lb{{\overline l}}
\def\yb{{\overline y}}

\def\ldel{{\overleftarrow{\del}}}
\def\rdel{{\overrightarrow{\del}}}
\def\ldeldel{{\overleftarrow{\del^2}}}
\def\rdeldel{{\overrightarrow{\del^2}}}
\def\ldelb{{\overleftarrow{\bar{\del}}}}
\def\rdelb{{\overrightarrow{\bar{\del}}}}

\def\ba{{\bf a}}
\def\bk{{\bf k}}
\def\bl{{\bf l}}
\def\bp{{\bf p}}
\def\bq{{\bf q}}
\def\br{{\bf r}}
\def\bt{{\bf t}}
\def\bu{{\bf u}}
\def\bv{{\bf v}}
\def\bx{{\bf x}}
\def\by{{\bf y}}
\def\bA{{\bf A}}
\def\bB{{\bf B}}
\def\bR{{\bf R}}
\def\bV{{\bf V}}

\def\bz{{\boldsymbol{\zeta}}}

\def\bone{{\bf 1}}

\def\va{{\vec a}}
\def\vk{{\vec k}}
\def\vp{{\vec p}}
\def\vq{{\vec q}}
\def\vx{{\vec x}}
\def\vy{{\vec y}}
\def\vu{{\vec u}}
\def\vv{{\vec v}}
\def \vH{{\vec H}}
\def \vg{{\vec g}}

\def\vs{{\vec \sigma}}
\def\vtau{{\vec \tau}}

\newcommand{\ov}[1]{\overrightarrow{#1}}

\def\frA{\mathfrak{A}}
\def\frB{\mathfrak{B}}
\def\frC{\mathfrak{C}}
\def\frD{\mathfrak{D}}
\def\frE{\mathfrak{E}}
\def\frF{\mathfrak{F}}
\def\frG{\mathfrak{G}}
\def\frH{\mathfrak{H}}
\def\frM{\mathfrak{M}}
\def\frN{\mathfrak{N}}
\def\frR{\mathfrak{R}}
\def\frW{\mathfrak{W}}

\def\fra{\mathfrak{a}}
\def\frb{\mathfrak{b}}
\def\frf{\mathfrak{f}}
\def\frg{\mathfrak{g}}
\def\frh{\mathfrak{h}}
\def\frl{\mathfrak{l}}
\def\frs{\mathfrak{s}}
\def\fri{\mathfrak{i}}
\def\frj{\mathfrak{j}}

\def\ma{\mathfrak{a}}
\def\mg{\mathfrak{g}}
\def\mh{\mathfrak{h}}
\def\mR{\mathfrak{R}}
\def\mN{\mathfrak{N}}

\newcommand{\nn}{{\nonumber}}

\def\d{\delta}\def\D{\Delta}\def\ddt{\dot\delta}

\def\pa{\partial} \def\del{\partial}
\def\xx{\times}
\def\uno{\mbox{1 \kern-.59em {\rm l}}}

\def\trp{^{\top}}
\def\inv{^{-1}}
\def\dag{\dagger}
\def\pr{^{\prime}}

\def\rar{\rightarrow}
\def\lar{\leftarrow}
\def\lrar{\leftrightarrow}

\newcommand{\0}{\,\!}      
\def\one{1\!\!1\,\,}
\def\im{\imath}
\def\jm{\jmath}

\newcommand{\tr}{\mbox{tr}}
\newcommand{\slsh}[1]{/ \!\!\!\! #1}

\def\vac{|0\rangle}
\def\lvac{\langle 0|}

\def\hlf{\frac{1}{2}}
\def\ove#1{\frac{1}{#1}}
\newcommand{\hot}[1]{\frac{#1}{2}}

\def\Box{\square}
\def\CC {\mathbb{C}}
\def\FF {\mathbb{F}}
\def\RR{\mathbb{R}}
\def\NN{\mathbb{N}}
\def\ZZ{\mathbb{Z}}
\def\bb#1{{\bf #1}}
\def\bcomment#1{}
\def\bfhat#1{{\bf \hat{#1}}}
\def\VEV#1{\left\langle #1\right\rangle}

\newcommand{\ex}[1]{{\rm e}^{#1}} \def\ii{{\rm i}}

\newcommand{\lrbrk}[1]{\left(#1\right)}
\newcommand{\lrsbrk}[1]{\left[#1\right]}
\newcommand{\sfrac}[2]{{\textstyle\frac{#1}{#2}}}

\def\stw{{\sqrt{2}}}

\def\rf {{\rm f}}
\def\ri {{\rm i}}
\def\rj {{\rm j}}
\def\rn {{\rm n}}
\def\rk {{\rm k}}
\def\rl {{\rm l}}
\def\rr {{\rm r}}
\def\rs {{\scriptscriptstyle \rm S}}
\def\rt {{\scriptscriptstyle \rm T}}

\def\rQ {{\scriptscriptstyle \rm \cQ}}
\def\rR {{\scriptscriptstyle \rm \cR}}

\def\cQb{{\cal \Qb}}
\def\cRb{{\cal \Rb}}
\def\cWb{{\cal \Wb}}

\def\fd {{\rm N}}
\def\afd {{\overline{\rm N}}}

\def \II {I\hspace{-.1em}I\hspace{.1em}}
\def \IIA {\mbox{\II A\hspace{.2em}}}
\def \IIB {\mbox{\II B\hspace{.2em}}}
\def \gs {g^s}
\def \ls {\lambda^s}

\def \I {{\cal I}}
\def \qs {q\hspace{-.53em}/\hspace{.15em}}
\def \ks {k\hspace{-.53em}/\hspace{.15em}}
\def \YM {{\mbox{\tiny YM}}}
\def \gym {g_{\YM}}

\def \Lc {\L_c}
\def\IR{\relax{\rm I\kern-.18em R}}
\def \id {{\bf 1}}

\def\cci{\ell}
\def\ccj{\ell'}

\def\bbq{\pmb{q}}

\newcommand{\para}[1]{\vskip 0.1cm {\noindent{#1}} \vskip 0.1cm}
\newcommand{\parabf}[1]{\vskip 0.3cm {\noindent{\bf #1}} \vskip 0.0cm}
\newcommand{\parait}[1]{\vskip 0.3cm {\noindent{\it #1}} }
\newcommand{\paratt}[1]{\vskip 0.1cm {\noindent{\tt #1}} \vskip 0.1cm}
\newcommand{\parasl}[1]{\vskip 0.1cm {\noindent{\sl #1}} \vskip 0.1cm}
\newcommand{\parasf}[1]{\vskip 0.1cm {\noindent{\sf #1}} \vskip 0.1cm}
\newcommand{\parasc}[1]{\vskip 0.1cm {\noindent{\sc #1}} \vskip 0.1cm}
\newcommand{\paraun}[1]{\vskip 0.1cm {\noindent\underline{\sf #1}}}

\begin{document}
\preprint{ NCTS-TH/2003}
\preprint{KOBE-COSMO-20-01}

\title{
  Gravitational Waves
  in Axion Dark Matter }

\author{Chong-Sun Chu$^{*,**}$}
\author{Jiro Soda$^{\natural}$}
\author{Daiske Yoshida$^{\natural}$}
\affiliation{$^*$Physics Division, National Center for Theoretical Sciences,
National Tsing-Hua University, Hsinchu, 30013, Taiwan\\
$^{**}$Department of Physics, National Tsing-Hua University, Hsinchu 30013, Taiwan
\\\
$^\natural$Department of Physics, Kobe University, Kobe 657-8501, Japan}
\date{\today}
\pacs{
	04.50.Kd	
}

\begin{abstract}
  Axion dark matter is interesting as it allows a natural coupling  to  the
  gravitational Chern-Simons term.
  In the presence of an axion background, the gravitational Chern-Simons
  term produces parity violating effects in the gravitational sector, in particular
  on the propagation of gravitational waves.
Previously, it has been shown that the coherent oscillation
of the axion field leads to a parametric amplification of
gravitational waves with a specific frequency.
In this paper we focus on the parity violating effects of the Chern-Simon
coupling and show the occurrence of  gravitational birefringence. We also find
deviation from the speed of light of the 
velocity of the gravitational
waves.  We  give
constraints on the axion-Chern-Simons coupling constant
and the abundance of the axion dark matter from the observation of
GW170817 and GRB170817A.
\end{abstract}

\maketitle

\section{Introduction}

The direct detection of gravitational waves in 2015 has
widen the frontiers in research in
fundamental physics~\cite{Abbott:2016blz}.
Indeed, we are
now
in the era of gravitational wave astrophysics and
multi-messenger astronomy
and it is now possible to
use  gravitational wave signals to
discriminate the various models beyond the standard model of particle
physics as well as of cosmology.  Moreover, the discovery of
gravitational waves GW170817~\cite{TheLIGOScientific:2017qsa} from a
neutron star binary
has sparked new interests in
the study of nuclear physics.
Remarkably, the observation of the optical counterpart of GW170817,
GRB170817A~\cite{Monitor:2017mdv}, has given a constraint on the
velocity of gravitational waves, which killed many modified theories
of gravity~\cite{Nishizawa:2017nef,Arai:2017hxj,Nishizawa:2018srh}.

In gravitational physics, there are three processes to be studied,
namely,
the production, propagation and detection of gravitational waves.
The production process and detection process have been well studied.
On the other hand, the propagation process has been
mostly
regarded as a
trivial problem.  In fact, in a Minkowski background, the
gravitational wave equation
is merely a conventional
scalar
wave equation.  Even in the presence of the
conventional matter, it is easy to solve propagation problem of
gravitational waves in the curved background.  However,
things can get more interesting with axions. As a pseudoscalar,
coupling to a gravitational Chern-Simons term is allowed
\cite{Campbell:1990fu,Lue:1998mq,Jackiw:2003pm}. As a result,
parity symmetry is broken in the presence of an axion background.
A peculiar
feature of the axion dark matter is coherent oscillations of the axion
field, which may affect the propagation of electromagnetic
waves~\cite{Yoshida:2017ehj} and gravitational
waves~\cite{Soda:2017sce, Yoshida:2017cjl}.

According to string theory, axions are ubiquitous in the
universe~\cite{Arvanitaki:2009fg}.  Remarkably, the mass of string
axions can take values in the broad range from $10^{-33}$ eV to
$10^{18}$ GeV.  In fact, it has been known that the axion is a natural
candidate of an inflaton and induces a circularly polarized
gravitational waves~\cite{Satoh:2007gn,Satoh:2008ck}.  Recently, an
axion has been intensively studied as a candidate for the dark matter.
As the dark matter, we can consider the axion with mass from
$10^{-23}$ eV to $10^{3}$ eV.  The lower bound comes from observations
of cosmic background radiations and the upper bound comes from
observations of X-ray backgrounds~\cite{Marsh:2015xka}.

In this paper, we comprehensively investigate gravitational waves
propagating in
an
axion background.  First of all, we review
the results of the previous work on the parametric resonance of gravitational waves
\cite{Soda:2017sce, Yoshida:2017cjl}.
Then, we focus on the gravitational birefringence and the velocity
modulation of gravitational waves.  In particular, we discuss
constraints on the Chern-Simons coupling and the abundance of the
axion dark matter from observations of the velocity of gravitational
waves.

The organization of the paper is as follows.  In section II, we
introduce basic equations
for gravitational waves in axion-Chern-Simons gravity. We also discuss a
potential
ghost mode and
a cutoff scale
that is needed in order to avoid the occurrence of this un-physical feature.
In section III, we analysis the propagation of gravitational waves in
a background of coherently oscillating axions.
In section IV, we review the parametric
amplification of gravitational waves and present consistency checks.
In section V, we consider the gravitational birefringence, which can
be regarded as a gravitational Faraday rotation.  In section VI, we
derive the velocity of gravitational waves. We discuss constraints on
the Chern-Simons coupling constant and the abundance of the axion dark
matter using the observation of the velocity of gravitational waves.
The final section is devoted to the conclusion.

\section{Gravitational Waves in Dynamical Chern-Simons Gravity}

Let us consider  the action of dynamical Chern-Simons gravity
\be\label{act0}
 S = \k\int{\rm d}^4x\sqrt{-g}R
 -\int{\rm d}^4x\sqrt{-g}\left[\frac{1}{2}\nabla^\mu\Phi\nabla_\mu\Phi
   +V(\Phi)\right]  + S_{{\rm dCS}},
 \ee
where the first term of action is the Einstein-Hilbert term,
$g$ is the determinant of the metric $g_{\mu\nu}$ and 
$\k=1/(16\pi G)$. The second term describes an action of an
 axion field $\Phi$.
In this paper we will take $\Phi$ as a dark matter candidate and exploit
existing observational constraints on its mass. 
The last term in \eq{act0}
\be
S_{{\rm dCS}}=
 \frac{1}{4}
 \int{\rm d}^4x\sqrt{-g} F (\Phi)R\tilde{R},\qquad
 R\tilde{R}:=\frac{1}{2}\epsilon^{\alpha\beta\gamma\delta}R_{\alpha\beta\rho\sigma}
 {R_{\gamma\delta}}^{\rho\sigma}  
 \label{act-dCS}
 \ee
is the dynamical Chern-Simons action with
$\epsilon^{\alpha\beta\gamma\delta}$ being the Levi-Civita tensor density.
We have allowed a nontrivial coupling $F(\Phi)$ of the
axion
field to
the $R\tilde{R}$ Chern-Simons term.
Otherwise, the dynamical Chern-Simons
term is topological and won't affect the equation
of motion.

We are interested in the effect of the dark matter Chern-Simons coupling
on the propagation of gravitational wave. Before we start, we need to fix
the background. Let us consider a background spacetime with  
  spatial isotropy and  homogeneity
  \be
  ds^2 = g_{\mu\nu}{\rm d}x^\mu{\rm d}x^\nu =
  - dt^2 + a^2(t) \delta_{ij}{\rm d}x^i{\rm d}x^j  \ .
  \ee
Due to its structure,
the Chern-Simons term does not contribute to the equation of motion of
the isotropic and homogeneous universe. As a result, we have the equations
of motion,
\begin{eqnarray}
  && 3H^2 = \frac{1}{2\k}
  \left(\frac{1}{2}\dot{\Phi}^2 + V(\Phi)\right)  \label{eq1} \\
 && \dot{H}+ 3 H^2
 =\frac{1}{2\k} V(\Phi) \label{eq2} \\ 
 && \ddot{\Phi} + 3 H \dot{\Phi}  +  V' (\Phi) =0 
 \label{eom},
\end{eqnarray}
where a dot denotes a time derivative and
$H= \dot{a}/a$ is the Hubble parameter.
Generally the dark matter has a mass
\be
V(\Phi) = \frac{1}{2} m^2 \Phi^2.
\ee
We will ignore self interaction as it is not relevant for our analysis.

\subsection{Action for  gravitational waves}

Let us now derive the quadratic action for the gravitational
waves from the action (\ref{act0}). 
The tensor perturbation reads
\be
 ds^2=g_{\mu\mu} dx^\mu  dx^\nu
= -dt^2
 +a^2(t) (\delta_{ij}+h_{ij}) dx^i dx^j \ ,
\ee
where  $h_{ij}$ satisfies the transverse-traceless conditions $h_{ij,j}=h_{ii}=0$.
Substituting the metric into \eq{act0}, 
we obtain the quadratic action
\be \label{S-quad}
  S =
\frac{\k}{4}
\int dt d^3x\;  a^3\left[  \dot{h}^{ij}\dot{h}_{ij} - \frac{1}{a^2}h^{ij|k}h_{ij|k}
  +\frac{\dot{F} }{\k a} \epsilon^{ijk}
  \left(\dot{h}_{ai}{\dot{h}^a}_{k|j}- \frac{1}{a^2}
  h_{ai|b}{{{h^a}_k}^{|b}}_{|j}\right)
       \right], 
\ee
where the stroke $|$ denotes a covariant derivative with respect to the spatial
coordinates. 
Here, we have used the convention $\epsilon^{0ijk}=-\epsilon^{ijk}$ with
$\epsilon^{123}\equiv 1$. It is convenient to expand $h_{ij}$
in terms of circular polarization basis
\begin{align}
 h_{ij}(\eta,{\bf x})=
 \sum_{A={\rm R, L}}\int\frac{{\rm d}^3k}{(2\pi)^3}h_A(t,{\bf k})
 {\rm e}^{i{\bf k}\cdot{\bf x}}p^A_{ij},
\end{align}
where $p^A_{ij}$  are the circular polarization tensor
defined by
\begin{align}
p^{\rm R}_{ij}\equiv\frac{1}{\sqrt{2}}
\left(p^{ +}_{ij}+ip^{\times}_{ij}\right),
\quad
p^{\rm L}_{ij}\equiv\frac{1}{\sqrt{2}}
\left(p^{ +}_{ij}-ip^{\times}_{ij}\right)
\end{align}
and the polarization tensors $ p^{ +}_{ij}$ and $p^{\times}_{ij}$
are the plus and cross modes respectively. 
The polarization tensors are normalized as 
\be
p^{*A}_{ij} p^B_{ij} =2 \delta^{AB}, \quad A,B ={\rm R,L}
\ee
and satisfies the helicity condition
\begin{align}
i \epsilon_i{}^{sj}\hat{k}_s p^A_{jk}=\rho_A p^A_{ik}
\quad \mbox{with}\quad \rho_{\rm R}= +1, \; \rho_{\rm L}= -1,
\end{align}
where $\hat{k^i} := k^i/k$ is the unit vector in the direction of
propagation of the gravitational wave.
The circular polarization modes are special as they diagonalize
the dynamical Chern-Simons action term and we obtain the quadratic action
\eq{S-quad}
\begin{eqnarray}
S = \frac{\k}{2} \sum_{A={\rm R,L}} \int dt \int \frac{d^3 k}{(2\pi)^3}\; a^3
B_A\left(
  \dot{h}_A{}^2 - \frac{k^2}{a^2} h_A{}^2
       \right)  \ .
\end{eqnarray}
where the function $B_A$ is defined by
\be
B_A(t,k):= 1- \frac{k \r_A}{\k}\frac{\dot{F}}{a}.
    \ee
In order for the perturbation to be stable, it is required that 
\be \label{ss}
B_A >0
\ee
 for both polarizations $A$.
One can derive from this
a natural upper bound to the energy scale of the gravitational
wave in the theory~\cite{Dyda:2012rj}.  
This is in contrast to the parity violating gravitational waves
in Lorentz violating gravity~\cite{Takahashi:2009wc}.

\section{Gravitational waves propagating in axion dark matter}

In this section, we suppose that a source in our cosmological horizon
emits gravitational waves propagating in the axion dark matter background. 
In this situation, we can neglect the effect of cosmic expansion
of the Universe in the dynamical equation \eq{eom} of the axion. 
To see this, let us note that
the Friedmann equation \eq{eq1} set the Hubble parameter to be of the order of
$H \sim \sqrt{\r/\k} \sim \sqrt{\r}/M_{\rm pl}$, where
$M_{\rm pl} := \sqrt{2 \k}$ is the reduced Planck mass.
On the other hand, the scalar field  changes at a rate
determined by the mass scale: $\dot{\Phi} \sim m$.
The cosmological expansion in \eq{eom} is negligible if
\be \label{cond-exp}
H/m \sim \sqrt{\r}/(m M_{\rm pl})  \ll 1.
\ee
Since an upper bound of  the dark matter
density is  given by
\be \label{rho-value}
\r = 0.3{\rm GeV/cm^{3}}=2.3\times10^{-6}{\rm eV^{4}} \ ,
\ee
the condition
\eq{cond-exp} is always  satisfied for the axion dark matter with
a mass $m>10^{-30}$ eV.

For simplicity, let us consider a linear dynamical Chern-Simon coupling
\be
F(\Phi) = \a \Phi
\ee
Following \cite{Yagi:2012vf}, we express
the Chern-Simon coupling constant $\a$ in terms of a length
$\ell$ as
\be
\a = \frac{M_{\rm pl}}{2} \ell^2,
\ee
where $M_{\rm pl} = \sqrt{2 \k}$ is the reduced Planck mass. 
The coupling constant
$\ell$ is experimentally constrained by the Gravity Probe B as~\cite{AliHaimoud:2011fw}
\be
\ell \leq 10^8 {\rm km}.
\ee

Now, we can write down the condition \eq{ss} in the present context. 
The axion satisfies the equation of motion
\be
\ddot{\Phi} + m^2 \Phi =0.
\ee
This can be solved as
\be\label{phi-t}
\Phi=\Phi_0 \cos(m t),
\ee
where, without loss of generality, we have made a choice of time so that
the phase in \eq{phi-t} is zero.
In this case, the no-ghost condition gives 
\be
k < k_{{\rm g}} := \frac{1}{\a m \Phi_0/\k} =\frac{\k}{\a \sqrt{2\r}}
\ee
where
\be
\r := \frac{1}{2}m^2 \Phi_0^2
\ee
is the energy density of the dark matter field.
Once we used the observed energy density,  
  the amplitude can be determined as
\be
\Phi_{0}\simeq2.1\times10^{7}{\rm eV}\,
\left(\frac{10^{-10}{\rm eV}}{m}\right)\sqrt{\frac{\rho}{0.3{\rm GeV/cm^{3}}}} \ .
\ee
Finally, we can deduce the cutoff frequency $f_{{\rm g}}$ as
\be
f_{{\rm g}}\equiv \frac{k_{{\rm g}}}{2\pi}= 1.1\times 10^9{\rm Hz}\left(\frac{10^{8}{\rm km}}{\ell}\right)^{2}\sqrt{\frac{0.3{\rm GeV/cm^{3}}}{\rho}} \ .
\ee
Above this scale, we cannot use the Chern-Simons gravity to describe
the propagation of gravitational waves in the axion dark matter. 

On energy scales below this,  the equation of motion of gravitational
waves in the Chern-Simons gravity is given by
\be
\ddot{h}_A+D_A \dot{h}_{A}+\frac{k^{2}}{a^{2}}h_{A}=0
\ee
where the function $D_A$ is defined by
\be
D_A (t,k):= 3H+
\frac{\dot{B}_A}{B_A} \ .
\ee
The first term is
due to  cosmological expansion, which we can ignore.
The second term is due to the 
dark matter background \eq{phi-t}.
It is convenient to introduce the dimensionless parameter
\be \label{d-def}
\d := m^2 \a \Phi_0 /\k
\ee
in terms of which we have 
\be
B_A = 1+ \frac{\r_A\d}{m} k \sin(mt).
\ee
The parameter can be estimated  as  
\be
 \d \simeq 2.3\times10^{-5}\left(\frac{m}{10^{-10}{\rm eV}}\right)\,
\left(\frac{\ell}{10^{8}{\rm km}}\right)^{2}
\sqrt{\frac{\rho}{0.3{\rm GeV}/{\rm cm}^{3}}}  \ .
\ee

In the following sections, 
we analyze the features of gravitational
waves in the axion dark matter background.

\section{Parametric Resonances}

Apparently, the dynamical Chern-Simons coupling
induces the parity violation in the presence of the axion dark matter.
In fact, the equations for each circular polarization modes are different.
Generically,  the polarization dependent effect
in gravitational waves characterized by $\d$ is small for typical
values of model parameters. 
However,  the effect can be exponentially enhanced
due to the resonance.

To see this, let us introduce a new variable $\Psi_A(t,k)$ defined by 
\be
h_{A}(t,k)=\exp\left(-\frac{1}{2}\int^t D_A(t',k) dt'\right)\Psi_{A}(t,k).
\ee
Then the equation of motion for the gravitational wave becomes
\be \label{Psi-eom}
\ddot{\Psi}_A + \omega_A^2 \Psi_A =0,
\ee
where
\be \label{om0}
\omega_A^2:= 
\frac{k^{2}}{a^{2}}-
\frac{1}{2}\dot{D}_A -\frac{1}{4} D_A^2.
\ee
To the leading order of  $\d$, the angular frequency is
given by
\be\label{om1}
\omega_A^2 = k^{2}\left(1+\rho_A f_{0}\sin\left(mt\right)\right)
\ee
and eq.\eq{Psi-eom}
takes the form of the Mathieu equation
\be
\ddot{\Psi}_A + k^{2}\left(1+\rho_A f_{0}\sin\left(mt\right)\right) \Psi_A =0,
\quad\mbox{where}
\quad f_{0} := \frac{1}{2}\frac{m}{k}\delta.
\ee
This describes an oscillator with a frequency $k$ 
pumped by the polarization dependent periodic force with
a magnitude $f_0$ and a frequency $m$. 
As is well known, the resonance occurs when
\be
k \sim m/2 \ . 
\ee
In this case, we obtain
\be
f_0 \sim \d \ .
\ee
The amplitude of gravitational waves $h_A$ grows exponentially
$ |h_A| \sim  e^{\Gamma t}$ with the growth rate given by
\bea
\Gamma &=& \frac{m \d}{8} \nn\\
&=&2.8\times10^{-16}{\rm eV}\,\left(\frac{m}{10^{-10}{\rm eV}}\right)^{2}\,
\left(\frac{\ell}{10^{8}{\rm km}}\right)^{2}\sqrt{\frac{\rho}
  {0.3{\rm GeV}/{\rm cm}^{3}}}.
\eea
We  can estimate the length $R_{\times10}$ which the gravitational
wave grows ten times bigger from the growth rate as follows,
\begin{align*}
R_{\times10} & =\frac{8}{m\delta}\\
& =5.2\times10^{-8}{\rm pc}\,\left(\frac{10^{-10}{\rm eV}}{m}
\right)^{2}\,\left(\frac{10^{8}{\rm km}}{\ell}\right)^{2}
\sqrt{\frac{0.3{\rm GeV}/{\rm cm}^{3}}{\rho}}.
\end{align*}
The range of the wave number for the resonance is given by
\[
\frac{m}{2}-\frac{m}{8}\delta\lesssim k\lesssim\frac{m}{2}+\frac{m}{8}\delta.
\]
This corresponds to a resonance width
$\Delta k_{\rm res}$:
\be \label{k-res}
\Delta k_{\rm res} =\frac{1}{4}m\delta.
\ee
The phenomenological consequence of this result has been discussed in
a previous paper \cite{Yoshida:2017cjl}. 
Recently, more serious comparison with gravitational observation are made in \cite{Sunghoon}

\subsection{Coherence length}

In the above analysis, we have
ignored the coherence issue of the dark matter background.
In principle, if the length scale of the gravitational perturbation becomes
comparable to the Jeans length scale, the dark matter cloud can
no longer remains as homogeneous and gravitational collapse will occurs.
In other words, the coherence can be sustained only within the Jeans scale.
As is known, the
Jeans length $r_{{\rm J}}$ of the axion dark matter can be deduced as
\bea
r_{{\rm J}} & =&6.7\times10^{20}{\rm
  eV}^{-1}\,\left(\frac{m}{10^{-10}\,{\rm
    eV}}\right)^{-\frac{1}{2}}\left(\frac{\rho}{0.3\,{\rm
    GeV/cm^{3}}}\right)^{-\frac{1}{4}}\nn \\
& =&4.3\times10^{-3}{\rm
  pc}\,\left(\frac{m}{10^{-10}\,{\rm
    eV}}\right)^{-\frac{1}{2}}\left(\frac{\rho}{0.3\,{\rm
    GeV/cm^{3}}}\right)^{-\frac{1}{4}}.
\eea
Thus, the condition $R_{\times10} \leq r_{\rm J}$ is necessary
for  the resonance to occur. This is satisfied for $m \geq m_c$
where the critical mass $m_c$ is given by
\be
m_{{\rm c}} =
5.3\times10^{-14}\,{\rm eV}\left(\frac{10^{8}{\rm km}}{\ell}
\right)^{\frac{4}{3}}\left(\frac{0.3{\rm GeV}/{\rm cm}^{3}}{\rho}
\right)^{\frac{1}{6}} \ .
\ee

\subsection{One more consistency check}

For having the resonance, we need one more condition.
Axion dark matter within the dark matter halo has a virial velocity.
From the simple dimensional analysis, one can estimate  the virial velocity as
\bea
v_{{\rm vir}}  =1.5\times 10^{-11}     \left(\frac{m}{10^{-10}\,{\rm
    eV}}\right)^{-\frac{1}{2}}\left(\frac{\rho}{0.3\,{\rm
    GeV/cm^{3}}}\right)^{\frac{1}{4}}\nn \ .
\eea
This velocity induces fluctuations
in the frequency of the axion dark matter given by
\be
\Delta k_{{\rm vir}}=\frac{1}{2}mv_{{\rm vir}}^{2}
\ee
If the fluctuations are
larger than the
band width
$\Delta k_{{\rm res}}$
of the parametric resonance, 
the amplitude
cannot grow efficiently.  
This is characterized by the ratio
\be
\gamma := \frac{\Delta k_{{\rm vir}}}{\Delta k_{{\rm res }}}.
\ee
If  $\gamma$ satisfies
\be
\gamma \ll 1,
\ee
the amplitude of gravitational waves grows. On the other hand, if 
$\gamma$ satisfies
\be
\gamma \geq 1,
\ee
then the
frequency of the axion dark matter easily escape from the resonance band 
 and  the gravitational wave never grows.
Since it is calculated as
\be
\gamma = 2.0\times  10^{-17} \left(\frac{m}{10^{-10}{\rm
    eV}}\right)^{-2}    \left(\frac{l}{10^{8}{\rm
    km}}\right)^{-2}\, ,
\ee
 we see that
 the amplitude of gravitational waves grows.

\section{Gravitational Faraday rotation}

The fact that the angular frequency \eq{om1}
is different for  different circular polarization modes implies that
the phase velocity 
\be
v_{A}^{(p)}:=
\frac{\omega_A}{k}=\sqrt{1+\r_A f_0 \sin\left(mt\right)}.
\ee
is different for  different circular polarization modes. 
It also implies a phase shift
between the R and L polarization arises as the wave propagates:
\be
\D \phi = \int_0^t dt' (\o_{\rm R} - \o_{\rm L}) = \frac{\d}{2} \cos(mt)
+ \cO(\d^2). 
\ee
This is characterized by a period of phase oscillation
\be
T \sim \left(\frac{m}{10^{-10}{\rm eV}}\right)^{-1} \sim 10^{-6} \,{\rm s} \ .
\ee
The  amplitude of the gravitational  Faraday rotation is given by
\be
(\D \phi)_{\rm max} = \frac{\d}{2}
=1.1\times10^{-5}\left(\frac{m}{10^{-10}{\rm eV}}\right)
\left(\frac{\ell}{10^{8}{\rm km}}\right)^{2}
\sqrt{\frac{\rho}{0.3{\rm GeV/cm^{3}}}} \, .
\ee
Hence, the gravitational Faraday rotation is sizable for
axion with
$m \gtrsim 10^{-5}$ eV. 

\section{Velocity of gravitational waves}

The propagation of gravitational waves is characterized by the
group velocity
$v^{(g)}_A:= \partial\omega_ A/\partial k$.
To the leading order of $\d$, we obtain
\be
v^{(g)}_A  =
1+\frac{1}{32}\frac{m^{2}}{k^{2}} \d^2 \sin^2(mt) + \cO(\d^3) \ ,
\ee
where we assumed $\delta\ll 1$ and $m\delta/k \ll 1$.
Note that it is independent of polarization up to
the second  order of $\d$. H
The group velocity is always greater than the speed of light
and has a maximal deviation of 
\begin{align}
  \Delta c & := \frac{1}{32}\frac{m^{2}}{k^{2}}\delta^{2} \nonumber \\
& =9.4\times10^{-7}\left(\frac{100{\rm Hz}}{f_{{\rm gw}}}\right)^{2}
  \left(\frac{\ell}{10^{8}{\rm km}}\right)^{4}\left(\frac{m}{10^{-10}{\rm eV}}
  \right)^{4}\left(\frac{\rho}{0.3{\rm GeV/cm}^{3}}\right).
\end{align}

The current upper bound of $\Delta c$
coming from the observation of GW170817 and GRB170817A is
\be
\Delta c\leq5\times10^{-16}.
\ee
 If we observe gravitational
waves oscillating in the low frequency and the axion dark matter which
have the heavier mass, we may give the stronger constraint on the
coupling constant $\ell$. Especially, if the gravitational waves through the core
of Galaxy, the velocity of the gravitational waves will be modified
strongly. 
For example, if we use  gravitational waves which have
the frequency about $1{\rm Hz}$ and we assume the density of the
axion dark matter is about $0.3{\rm GeV/cm^{3}}$ and
the mass of the axion dark matter is about $10^{-10}{\rm eV}$, 
the constraint on the Chern-Simons coupling constant reads
\be
\ell\leq 4.8\times10^{4}{\rm km} \ .
\ee
Once we obtain this constraint, by observing at a lower frequency, say $10^{-4}$Hz,
we can further constrain the Chern-Simons coupling constant as
\be
\ell\leq 4.8\times10^{2}{\rm km} \ .
\ee
For the extreme case of $f_{\rm gw} = 10^{-9}$Hz and $m=10^3$eV, we obtain the stringent constraint on the coupling constant
\be
\ell\leq 1.5 \times10^{-13}{\rm km} \ .
\ee
On the other hand, assuming the coupling constant $\ell =10^2$ km and
$f_{\rm gw} =10^{-9}$ Hz, we obtain the constraint on the abundance of the axion with $m=10^{-10} $ eV as
 \begin{eqnarray}
    \Omega_{\rm axion }    <     3.4\times 10^{-3}    \ .
\end{eqnarray}

 Thus, we see the gravitational waves
can provide useful constraints to the Chern-Simons coupling constant and the abundance of the axion dark matter.

\section{Conclusion}

We studied gravitational waves propagating in the axion dark matter.
In the presence of the axion, it is natural to consider the coupling of the axion to
the gravitational Chern-Simons term.  Since the axion condensation violates
the parity symmetry, there is a chance to observe
parity violation effects
in the gravity sector using gravitational waves. 
We found that the coherent oscillation of the axion field 
leads to the parametric amplification of gravitational waves with a
specific frequency. 
We investigated the gravitational birefringence induced by the difference
in the phase velocity of the different polarization modes.
We also derived the group velocity of gravitational waves which is
independent on the polarization at the leading order. 
In particular, we have given a constraint on the Chern-Simons coupling constant
and the abundance of the axion dark matter from the observation of 
GW170817 and GRB170817A.

There are various ways to proceed. It is important to perform a comparison
of  our result with
real data.
It is
interesting to  study gravitational wave propagating in  the
ultralight vector  dark matter~\cite{Nakayama:2019rhg,Nomura:2019cvc}.
It is
also
possible to extend the analysis to other higher spin  dark matter.

\vskip7mm
\section*{Acknowledgments}
J.~S. would like to thank Sunghoon Jung for useful comments and TaeHun Kim and Yuko Urakawa
 for fruitful discussions. 
C.S.C. was supported in part by 
NCTS and the grant MOST  107-2119-M-007-014-MY3 
of the Ministry of Science and Technology of Taiwan.
J.~S. was
supported
in part
by JSPS KAKENHI
Grant Numbers JP17H02894, JP17K18778, JP15H05895, JP17H06359, JP18H04589.
J.~S. was also supported by JSPS Bilateral Joint Research
Projects (JSPS-NRF collaboration) `` String Axion Cosmology.''
Discussions during the YITP workshop YITP-T-19-02 on ``Resonant instabilities in cosmology'' were useful for this work. 
D.~Y. was supported by Grant-in-Aid for JSPS Research Fellow and JSPS KAKENHI Grant Numbers 17J00490.




\end{document}